\documentclass[12pt,twocolumn,english]{article}
\usepackage[T1]{fontenc}
\usepackage[latin9]{luainputenc}
\usepackage{geometry}
\usepackage{babel}
\usepackage{float}
\usepackage{calc}
\usepackage{graphicx}
\usepackage{setspace}
\PassOptionsToPackage{normalem}{ulem}
\usepackage{ulem}
\usepackage[unicode=true]
 {hyperref}

\makeatletter

\newcommand{\noun}[1]{\textsc{#1}}
\providecommand{\tabularnewline}{\\}

\newcommand{\lyxrightaddress}[1]{
	\par {\raggedleft \begin{tabular}{l}\ignorespaces
	#1
	\end{tabular}
	\vspace{1.4em}
	\par}
}

\makeatother

\begin{document}
\title{Concerns about ground based astronomical observations: \noun{}\\
\noun{A step to Safeguard the Astronomical Sky} }
\author{Stefano Gallozzi\textsuperscript{(1)}, Marco Scardia\textsuperscript{(2)},
Michele Maris\textsuperscript{(3)}}
\maketitle

\lyxrightaddress{\textit{\footnotesize{}(1) stefano.gallozzi@inaf.it, }\\
\textit{\footnotesize{}$\quad$INAF--Osservatorio Astronomico di
Roma }\\
\textit{\footnotesize{}(2) marco.scardia@inaf.it, }\\
\textit{\footnotesize{}$\quad$INAF--Osservatorio Astronomico di
Brera }\\
\textit{\footnotesize{}(3) michele.maris@inaf.it, }\\
\textit{\footnotesize{}$\quad$INAF--Osservatorio Astronomico di
Trieste }\\
}
\begin{abstract}
This article aims to highlight the impact for ground based astronomical
observations in different windows of the electromagnetic spectrum
coming from the deployment of fleets of telecommunications satellites.
\\
A particular attention is given to the problem of crowding of circumterrestrial
space by medium/small size orbiting objects. Depending on their altitude
and surface reflectivity, their contribution to the sky brightness
is not negligible for professional ground based observations. With
the huge amount of about 50,000 new artificial satellites for telecommunications
planned to be launched in Medium and Low Earth Orbit, the mean density
of artificial objects will be of >1 satellite for square sky degree;
this will inevitably harm professional astronomical images. \\
Only one of these project, Starlink@SpaceX's, was authorized by US
Federal Communication Commission, F.C.C. and plans to deploy about
42,000 not geostationary satellites, which will shine from the 3rd
to the 7th magnitude in sky after sunset and before sun dawn. \\
All satellites will leave several dangerous trails in astronomic images
and will be particularly negative for scientific large area images
used to search for Near Earth Objects, predicting and, eventually,
avoiding possible impact events. Serious concerns are common also
to other wavelengths eligible for ground based investigation, in particular
for radio-astronomy, whose detectors are already saturated by the
ubiquitous irradiation of satellites communication from Space stations
as well as from the ground. Not to exclude the significant increase
in the risk of hitting into the \textquotedbl Kessler syndrome\textquotedbl{}
scenario with the deployment of all of these satellites.\\
Understanding the risk for astronomical community, a set of actions
are proposed in this paper to mitigate and contain the most dangerous
effects arising from such changes in the population of small satellites.
A dedicate strategy for urgent intervention to safeguard and protect
each astronomical band observable from the ground is outlined.\\
\uline{Keywords: }Satellites constellations, ground based astronomical
observations, ground based radio astronomy.
\end{abstract}

\section{Introduction}

The deployment of large fleets of small satellites planned or ongoing
for the next generation of global telecommunication networks can severely
harm ground based astronomical observations.

For centuries ground based astronomical observations have led to exceptional
progresses in our scientific understanding of the Laws of Nature.
Currently, the capability of ground based astronomical instrumentation
is endangered by the deployment of satellite fleets of unprecedented
size.

Astronomers all over the world are concerned for the sky coverage
and light/radio pollution produced by artificial satellites, which
represent a dramatic degradation of the scientific content for a huge
set of astronomical observations.

The same concerns have been expressed by the \textbf{International
Astronomical Union, IAU} {[}1{]} and other institutions since the
sky degradation is not only due to light pollution in the sky near
cities and the most populated areas, but it is also due to large artificial
satellite fleets crossing the sky producing bright parallel streaks/trails
at all latitudes.

This paper is organized as follow: section 2 describes the substantial
and complementary characteristic between ground based astronomical
facilities and those in orbit; satellite constellations are introduced
in section 3; section 4 illustrates how such constellations may affect
ground based observations; section 5 is devoted to discuss possible
mitigations; conclusions are in section 6.

\section{Ground Based and Space Based Astronomy }

The advent of space based astronomy (i.e. UVES, HST, Spitzer, Herschel,
Planck to cite just some of well known space telescopes) did not dismiss
ground based astronomy. On the contrary ground based astronomy and
space based astronomy complement each other. Without ground based
observations most of current space based astronomy would be useless
or impossible. The reasons for this statement are shortly outlined
below.

The effort of developing, deploying and operating a space based telescope
is bigger, even of more than an order of magnitude, than the effort
required for a ground based telescope of similar size (quantified
by the diameter of its main mirror). As an example it took more than
twenty years to develop the next large space telescope, the JWST planned
for launch in 2021, with a 6.5 meters mirror at the cost of the order
of ten billions euros. In the meantime the VLT, with four giant 8.2
m telescopes, two Magellan telescopes with 6.2 meters mirrors, and
other telescopes of similar size entered in service. The next generation
of new giant ground based telescopes as the GMT with a 25.5m mirror,
the ESO/eELT with a 39m mirror is expected to be putted online in
about half the time required for the JSWT and with a cost of about
one billion euros per telescope. 

As the ability of a telescope to reveal small, weak, far objects increases
with the area of its mirror which collects light, the sensitivity
of a telescope increases with the square of its diameter. So in ideal
conditions, the eELT will be able to collect thirtysix times the amount
of light collected by the JWST. 

Compared to a ground based telescope, a space based telescope has
the strong advantage of observing the sky outside the Earth atmosphere.
In some bands, as an example the X rays, the Far UV or the far IR,
the atmosphere blocks completely the incoming radiation. So that a
ground based telescope in those bands would be completely blind or
severely limited, but this is not true in other bands. As an example:
in visible light our atmosphere is quite transparent, so that to observe
in visible light from space is of little advantage with respect to
the case of observing from ground, save for the disturbance introduced
by atmospheric turbulence. However this last problem has been largely
mitigated in the last decades through a careful selection of the sites
where telescopes are installed, and by the adoption of technological
innovations such as the adaptive optics. 

A major limitation of space based telescopes is that they can not
be maintained, refurbished or repaired after launch. In this respect
HST has been a \textquoteleft unicum\textquoteright{} that hardly
will be possible to repeat. The obvious consequence is that the operational
lifetime of such telescopes is limited by the amount of consumables
which can be stored onboard (example coolant for the instruments)
and by the resilience of the instruments to the degradation induced
by ageing, radiations, micro meteoroids and so on. Compared to ground
based observatories, the average life-time of space based telescopes
is of the order of a couple of decades or less. On the contrary ground
based observatories lasts for several decades, with telescopes installed
at the beginning of the space era again working in a profitable manner.
Being impossible to replace its components, space based telescopes
suffer technological obsolescence when compared to ground based observatories.
In general space based telescopes are too expensive to be used to
validate new observing technologies, which usually are developed on
ground based telescopes first. 

There are strong limitations on the mass, the size, the technology
which can be sent in space, putting severe constraints on designing
and operating space telescopes. On the contrary, we are virtually
allowed to plan as large a ground telescope as we want, provided it
will not break under its own weight, it will be of practical use,
and its cost will be sustainable. 

Before to conclude it is important to stress that the arguments presented
above have not to be misused or misinterpreted as arguments against
space based astronomy. On the contrary in the modern professional
practice both kinds of astronomy are equally important, so that astronomers
involved in ground based astronomy are often involved in space based
astronomy and vice-versa. Our aim is to disprove the quite common
misunderstanding that ground based astronomy can be dismissed because
of the advent of space based astronomy.

\section{What are Satellites' Constellations}

Over the past decades, considerable effort has gone into designing,
building, and deploying satellites for many important purposes. Recently
networks, known as satellite constellations, have been deployed and
are planned in ever greater numbers mainly in low-Earth orbits for
a variety of purposes, including providing communication services
to underserved remote areas. Until 2019, the number of such satellites
was below 200, but that number is now increasing rapidly, with plans
to deploy potentially tens of thousands of them. If no action will
be put in place, several problems will soon arise in Astronomical
observations.

\subsection{The Iridium Satellite Constellation}

The Iridium satellite constellation provides \uline{L-band} voice
and data information coverage to satellite phones, pagers and integrated
transceivers over the entire Earth surface. The band used to provide
communication services are proper of LTE-Advanced and UMTS/HSDPA services
and operates from 1452 to 1492 MHz.\footnote{In Europe, the Electronic Communications Committee (ECC) of the European
Conference of Postal and Telecommunications Administrations (CEPT)
has harmonized part of the L band (1452--1492 MHz), allowing individual
countries to adopt this spectrum for terrestrial mobile/fixed communications
networks supplemental down-link (MFCN SDL). By means of carrier aggregation,
an LTE-Advanced or UMTS/HSDPA base station could use this spectrum
to provide additional bandwidth for communications from the base station
to the mobile device; i.e., in the downlink direction.} 

The constellation consists of \textbf{66 }active satellites in orbit,
required for global coverage, and additional spare satellites to serve
in case of failure. Satellites are in Low Earth Orbit, LEO at a height
of approximately \uline{781km} and inclination of $86.4{^\circ}$.
Orbital velocity of the satellites is approximately 27,000km/h. Satellites
communicate with neighboring satellites via Ka band inter-satellite
links. Each satellite can have four inter-satellite links: one each
to neighbors foreground and afterground in the same orbital plane,
and one each to satellites in neighboring planes to either side. These
satellites will cover an entire orbit around the Earth within roughly
\uline{100 minutes}.

\subsection{The SpaceX Starlink Constellation}

The US company SpaceX plans to put in orbit a very huge constellation
of 42,000 satellites, called Starlink. This constellation is aimed
to provide internet access. These satellites work in conjunction with
ground trans-receiver stations. A small set of Starlink satellites
is planned to be dedicated to military airforce {[}26{]} and/or research
purposes. This satellites fleet will be displaced in three orbital
shells: 
\begin{enumerate}
\item about \textbf{10,000} satellites at \uline{1150-kilometers-altitude}
orbit shell using the \uline{Ku band} (from 12 to 18 GHz) and \uline{Ka
bands} (from 26.5 to 40 GHz).
\item about \textbf{6,000} satellites in a \uline{550-kilometer-altitude}
orbit shell, using the same \uline{Ku} and \uline{Ka bands}. 
\item about \textbf{26,000} satellites in a \uline{340-kilometer-altitude}
orbit shell using the \uline{V-band} (from 40 to 75 GHz).\footnote{The 5th generation mobile networks (28, 38, and 60 GHz) will also
partially overlap with Ka and V bands. The V band at 60 GHz was used
by the world's first cross-link communication between satellites in
a constellation between the U.S. Milstar 1 and Milstar 2 military
satellites. The 60GHz frequency band is attractive to secure satellite
crosslinks because it allows high data rates, narrow beams and, lying
in a strong absorption band of oxygen, provides protection against
intercept by ground-based adversaries.}
\end{enumerate}
The first phase of deployment will put in orbit first the 550km-altitude
satellites, then those at 1150km-altitude. To the first phase a second
phase will follow with the deployment in orbit of the remaining inner
satellites at 340km-altitude. 

\begin{figure*}[t]
\includegraphics[width=0.8\paperwidth]{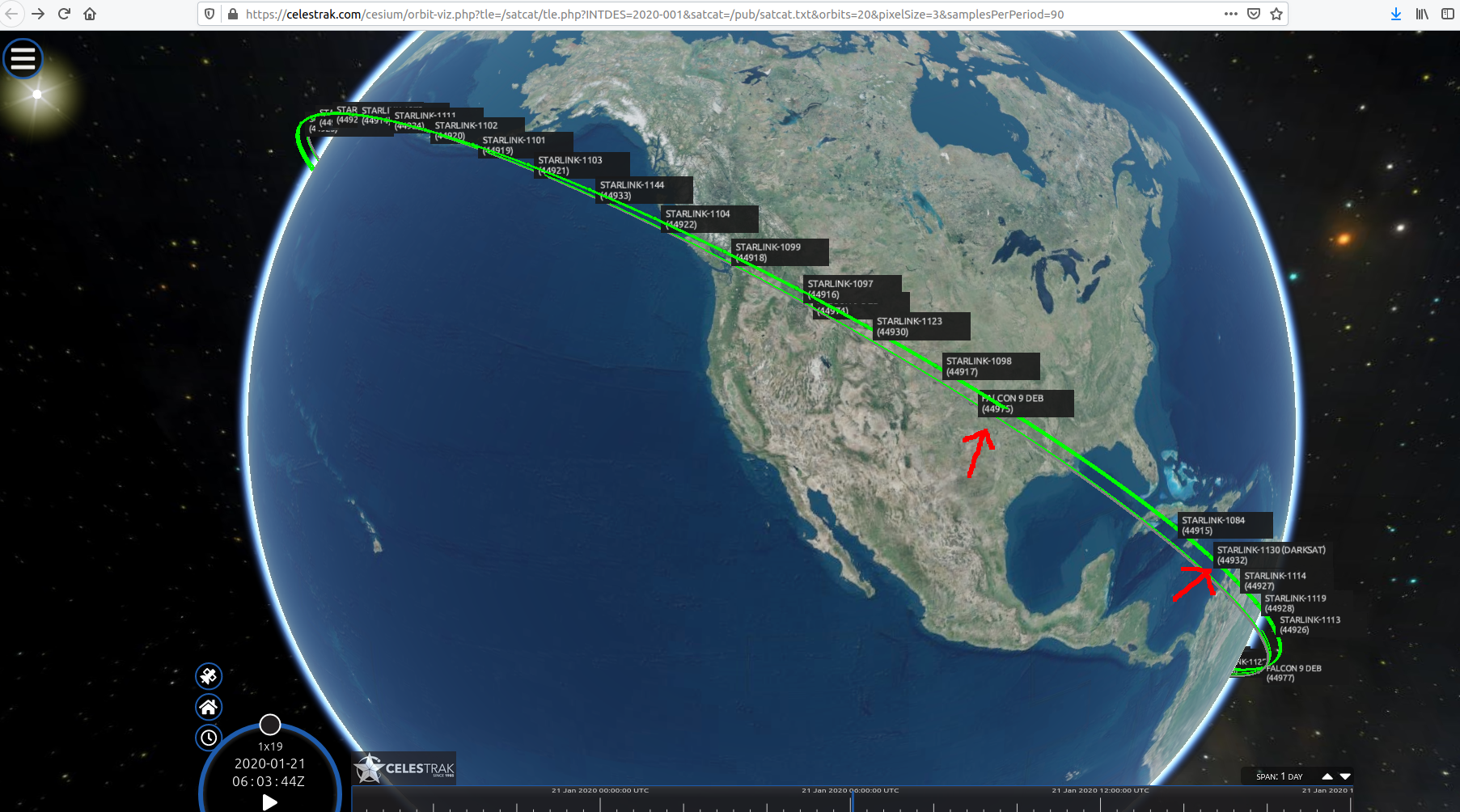}

\caption{First orbital planes of Starlink satellites after the 21 January 2020
launch of 60 satellites by the Falcon-X rocket to reach the total
number of orbiting satellites of 240. Starlink@SpaceX systems will
be turned on at 420 satellites, while the first broad band internet
service will be provided once \textasciitilde 1000 satellites will
be deployed. The red arrows indicates the Falcon9 orbiting bucket
and the experimental Starlink DARKSAT, with an experimental coating
to make it less reflective, and thus impact ground-based astronomical
observations less. }
\end{figure*}

Because the Starlink satellites can autonomously change their orbits,
observations cannot be scheduled to avoid them. 

\subsection{The OneWeb Satellite Constellation}

Is a UK project to provide global satellite Internet broadband services
to people everywhere and it is composed of a constellation of about
\textbf{5260 satellite} displaced in a circular LEO orbit shell of
about \uline{1,200km-altitude}. It will transmit and receive in
the \uline{Ku band}. 

There are about 6 testing satellites in orbit; in February 2020 other
34 satellites are planned to be launched from the Baikonur Site. Other
launches are scheduled in 2020.

\subsection{Other Constellations}

In the next years there could be \uline{over 50,000 new satellites
encircling the Earth} (at different altitudes) for various telecommunication
purposes but mainly delivering broad band internet from Space and,
considering the closeness to the Earth, internet signals will be provided
fast and with very low-latency. Table 1 is a non-exhaustive list of
principal satellite-constellation.
\begin{table*}[t]
\begin{tabular}{|c|c|c|c|c|}
\hline 
\textbf{Constellation Name} & \textbf{n. Satellites} & \textbf{Altitude {[}km{]}} & \textbf{Bands} & \textbf{Serv.Start}\tabularnewline
\hline 
\hline 
SpaceX - Starlink (USA) & 42,000 & 1150, 550, 340 & Ku, Ka, V & 2020\tabularnewline
\hline 
OneWeb (UK) & 5,260 & 1200 & Ku & 2020\tabularnewline
\hline 
Telesat (CAN) & 512 & \textasciitilde 1000 & Ka & 2022\tabularnewline
\hline 
Amazon - Kuiper (USA) & 3236 & 590, 630, 610 & ? & 2021\tabularnewline
\hline 
Lynk (USA) & thousands &  & ? & 2023\tabularnewline
\hline 
Facebook (USA) & thousands & 500-550 & ? & 2021\tabularnewline
\hline 
Roscosmos (RU) & 640 & 870 & ? & 2022-2026\tabularnewline
\hline 
Aerospace Sci.Corp. (CHI) & 156 & \textasciitilde 1000 & ? & 2022\tabularnewline
\hline 
\end{tabular}

\caption{Satellite-Constellation projects comparison.}
\end{table*}

A so large number of new self-drivng satellites in different low-altitude
orbiting shells could also impact on the capability to send in the
outer space new science related missions, since it would be impossible
to exactly predict the single positioning of each constellation satellite,
so that impact risk will strongly increase during scientific mission
launches.

\section{The impact of large satellite constellations on ground based astronomy}

The ground-based observatories with their large optical telescope
currently working (Very Large Telescope, VLT {[}2{]}, Large Binocular
Telescope, LBT {[}3{]}, ...) and those in construction (e.g. the Extremely
Large Telescope, ELT {[}4{]}, with a main 39-meter diameter mirror)
are essential complements to astronomical satellites, which are not
affected by reflections of the satellites. As introduced in the second
section, for astronomical satellites, missions costs and limitations
on size/weight preclude the launch of particularly large telescopes,
thus the difficulty in repairing and maintaining telescopes in space
means that \uline{the newest, most revolutionary technologies are
implemented only on ground-based telescopes}. Ground based telescopes
are fundamental to astronomy and the international community and single
states have invested for these ground based projects in past years
several tens of billions of dollars/euro trillions of dollars, therefore
they are requested to produce high rate of scientific results to repay
the initial public investment received.

What is really damaging such scientific results is the \uline{sky
degradation}. This is not only due to sky-glow / light pollution in
the sky near cities and the most populated areas (see Fig. 2), but
it is also due to artificial satellite fleets crossing and scarring
observations with bright parallel streaks/trails at all latitudes.

Astronomers are extremely concerned by the possibility that sky seen
from Earth may be blanketed by tens of thousands of satellites, which
will greatly outnumber the approximately 9,000 stars that are visible
to the unaided human eye. This is not some distant threat: it is already
happening. As seen in section 3, the US private company SpaceX has
already put 240 of these small satellites, collectively called Starlink,
in the sky and plans to constellate the whole sky with about 42,000
satellites. Thus, together with other telecommunication space projects
in the near future (see $\mathsection1.1$), in a very short term
there could be over 50,000 small satellites encircling the Earth (at
different altitudes) each of them damaging astronomical observations.
\begin{figure*}
\includegraphics[width=0.8\paperwidth]{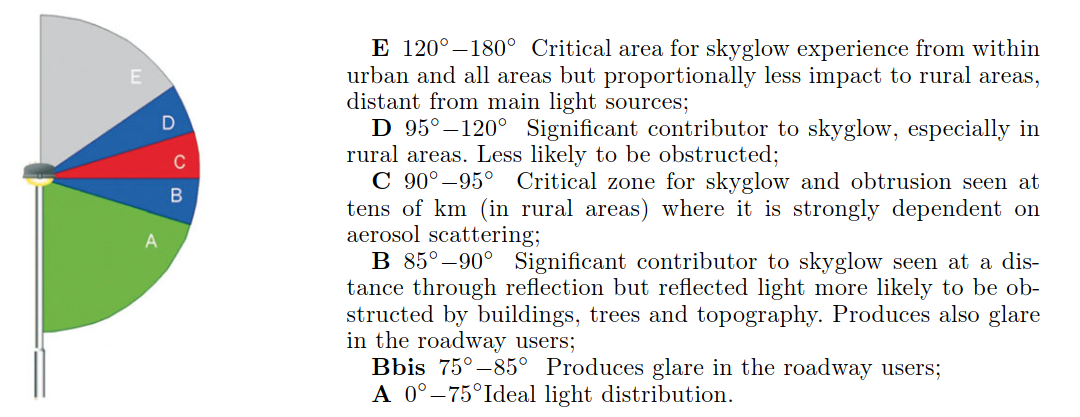}

\caption{Effect of Sky-glow and cut-off angle, showing the relative impact
of a luminaire's output contribution to skyglow. Picture taken from
``Starlight: a Common Heritage'', Cipriano Marin, IAC - ESP}
\end{figure*}

The closeness of satellites in LEO makes them more visible, and brighter
in the night sky especially when lighted by the Sun (e.g. satellites
launched by SpaceX are brighter than 99 percent of the population
of objects visible by the Earth orbit ).

As comparison, the current \uline{total number of cataloged objects
in Earth orbit} is less than 20,000 among spacecrafts, rocket bodies,
fragmented mission and other related debrids, so with only the nominal
Starlink fleet the total number of orbiting objects will increase
of 300\% (see Fig. 2 and 3).\footnote{It has been estimated that here are over 500,000 non-cataloged pieces
of space debris from the size of one square cm (or larger) orbiting
the Earth traveling up to 17,500 mph; millions of others are untraceable,
in addition to the around 20,000 cataloged objects in Earth orbit.
In 1978 this scenario was predicted by D.J Kessler, {[}30{]}, warning
governments about the devastating cascading effect of collision-induced
debris creation (the so-called \uline{Kessler syndrome}). The increase
number of LEO satellites makes the creation of a dense debrids network
belt around the Earth a possible scenario with devastating consequences
for the future of space exploration and telecommunications too. }

\begin{figure*}[t]
\includegraphics[width=0.8\paperwidth]{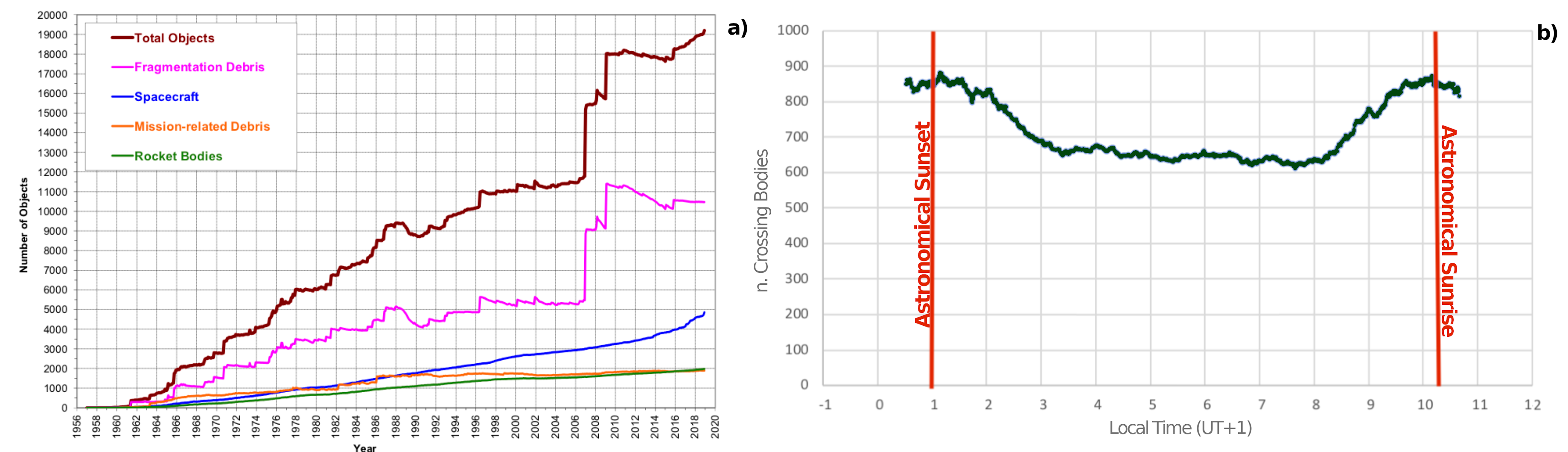}

\caption{a) Number of object in the Earth's orbit and b) Number of artificial
crossing bodies during an observing night.}
\end{figure*}

In the mid and long term, this will severely diminish our view of
the Universe, create more space debris and deprive humanity of an
unblemished view of the night sky. 

It has been considered that most of these satellites will be visible
to the naked eye (with a brightness between the 3rd and 7th magnitude)
particularly in the time after sunset and before sunrise. Consequently
they will reach the brightness of the stars in the Ursa Minor constellation.
There are only 172 stars in the whole sky exceeding the expected brightness
of Starlink satellites (see Tab. 2). Higher altitude LEO satellites
(e.g. over 1000km-altitude) will be visible all the night reaching
approximately the 8th magnitude.

\begin{table}[H]
\includegraphics[width=0.4\paperwidth,height=0.45\paperwidth]{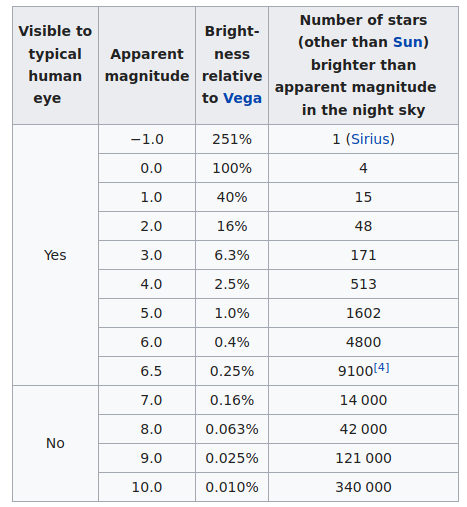}

\caption{Stars Apparent Magnitudes comparisons.}
\end{table}

The most important contribution on pollution of astronomical images
comes from the satellites in the higher orbits since the light directly
reflected by the Sun make them brighter during the night, instead
lower altitude satellites are foreseen to contribute negatively only
few hours after sunset and before sun dawn. It is possible to predict
the range of variability of apparent magnitudes of the LEO satellites
depending on the position and the altitude considering a mean density
of about 1 satellite per square degree (see Fig. 4a and 4b and {[}12{]}
for a whole sky simulation of 12,000 Starlink satellites at three
different altitudes). 

\begin{figure*}[t]
\includegraphics[width=0.8\paperwidth]{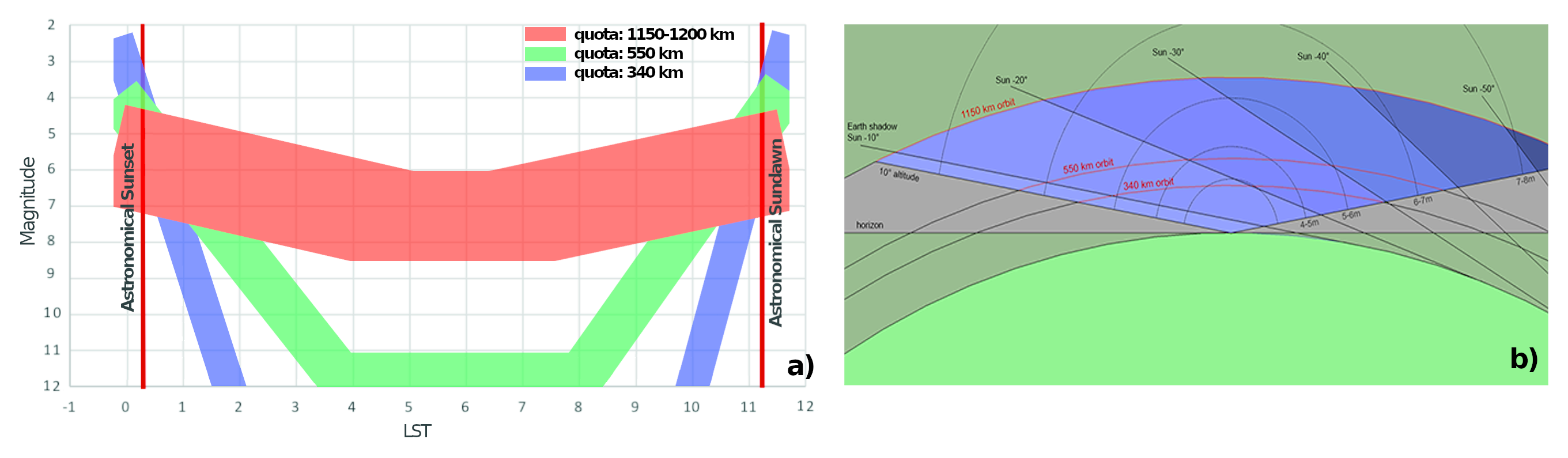}

\caption{a) Apparent magnitude of satellites during an observing night depending
on the altitude; b) Illumination factor depending on the Sun altitude
of the three orbital shells for Starlink satellites. For OneWeb satellites
the expected illumination fraction will be the same of highest starlink
orbital shell, see {[}28{]}.}
\end{figure*}

In Fig. 4b, illustrates how starlink orbital shells (shown in red)
are illuminated by the Sun when it is at different altitudes below
the horizon. We can see that the lower the sun is only the more distant
satellites will be illuminated. At certain stages the lowest shells
won't be visible at all, but the higher shells will be visible in
the northern part of the sky. Also the swarm of the satellites near
the horizon will be mostly invisible due to their distance and atmospheric
effects. It should be noted that the \textquotedbl worst case\textquotedbl{}
will be experienced \uline{during the summer, in northern half
of the sky, in the northern hemisphere, where the satellites will
be visible during the entire night}, though their brightness will
probably be lower than the bright overhead passes after the sunset
and before the sunrise. The \textquotedbl best case\textquotedbl{}
will be during the winter, at midnight, when the sky will be virtually
free of any satellites, except for the horizon; for details see {[}28{]}.

Thus with 50k satellites the \textquotedbl normality\textquotedbl{}
will be a sky crowded with artificial objects: every square degree
of the sky will have a satellite crawling in it along the whole observing
night accessible and visible by astronomical cameras and not only
by professional instrumentation.

It should also be noted that during nominal service operations SpaceX
expects to dismiss and replace from 2,000 to 8,000 Starlink satellites
every year, disintegrating them in the lower atmosphere, with all
related issues, see {[}27{]} page 4-5 for details.

\subsection{Impacts on ground based optical astronomy}

Wide-field survey telescopes will be particularly damaged, by the
presence of multiple saturated trails within camera images:
\begin{itemize}
\item LSST {[}5{]}, capable to scan and perform a survey of the entire sky
in three nights 
\item VST {[}6{]}, with its 268MegaPixels camera and a FOV of 1 square degree 
\item Pan-STARRS {[}7{]}, with its FOV of 7 square degrees and 1.4 Giga
pixels camera
\end{itemize}
Also deep/long exposures with small-field facilities will be unavoidably
impaired, see Fig. 5 and {[}12{]}. 
\begin{figure*}[t]
\includegraphics[width=0.8\paperwidth,height=0.6\paperwidth]{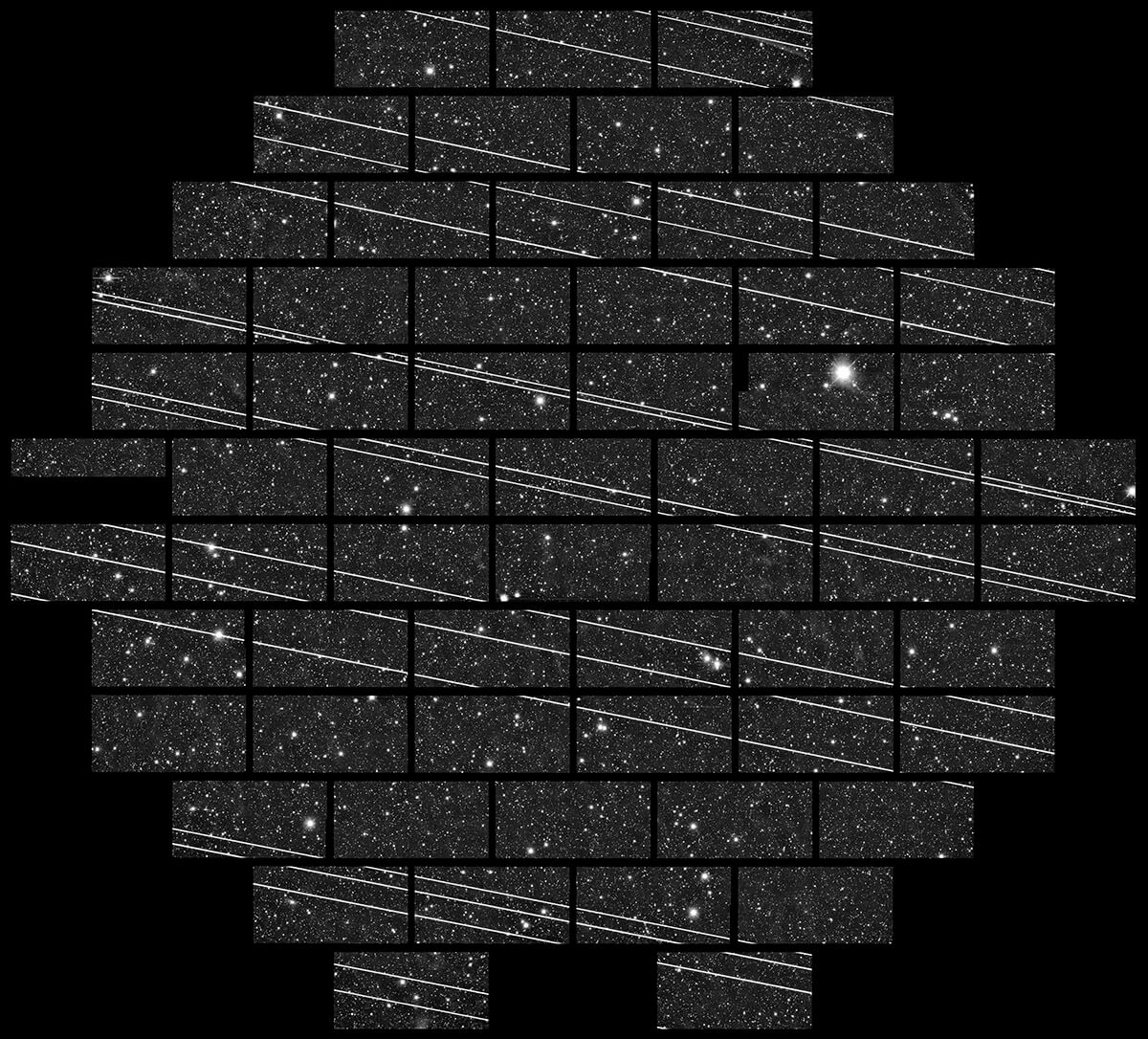}

\caption{Few Starlink satellites visible in a mosaic of an astronomical image
(courtesy of NSF\textquoteright s National Optical-Infrared Astronomy
Research Laboratory/NSF/AURA/CTIO/DELVE)}
\end{figure*}

This have also a dramatic impact on our safety because large area
astronomical observations and sky surveys are commonly used in search
for Near Earth Objects, NEOs, asteroids monitoring and other related
searches to guard the Earth from potential impact events: such \uline{satellite
constellations could negatively impact on the ability to prevent and
warn the whole humankind}.

The light pollution is extremely damaging for astronomical observations
at all wavelengths. To minimize the quantity of light reflected by
LEO satellites, Starlink has put in orbit an experimental version
of the Starlink satellites ( Starlink satellite n.1130 DARKSAT, see
Fig. n.1) making use of a non-reflecting paint on the body. It is
not clear how this coating will reduce the brightness of the satellite
since it is not possible to cover solar panels, which represents 75\%
of the satellite reflecting surface.

If the satellite body will be inhibited to reflect the sun light,
it will absorb radiation warming too much with possibile failures,
thus will probably increase the risk management for the whole fleet
and make the dark-coating solution ineffective or even counterproductive.

Moreover even if the brightness of the experimental satellite would
be below the naked eye sensitivity, astronomical images will continue
to see them (with resulting damage to their scientific content). 

Thus degradation for scientific observations will remain high also
with coating for three different reasons: 
\begin{enumerate}
\item Astronomical deep field camera images will continue to have trails
in long exposures depending on the filter limiting magnitude. 
\item Astronomical objects in the sky will be eclipsed, this will probably
harm time-dependent (variability) studies.
\item The reflectivity of a surface depends on the observational wavelength,
so what becomes dark in one part of the spectrum (e.g. visible), will
remains bright in other parts of the spectrum (e.g. infrared or radio).
\end{enumerate}

\subsection{Impacts on ground based radio-astronomy }

Even with best coating and mitigation procedures to decrease the impacto
on visual astronomical observations, what it is often omitted or forgotten
is that telecommunication constellations will shine in the radio wavelengths
bands, observable from the ground. 

The scientific needs of radio astronomers and other users of the passive
services for the allocation of frequencies were first stated at the
World Administrative Radio Conference held in 1959 (WARC-59). At that
time, the general pattern of a frequency-allocation scheme was:
\begin{enumerate}
\item that the science of radio astronomy should be recognized as a service
in the Radio Regulations of the \textbf{International Telecommunication
Union (ITU)}; 
\item that \textbf{a series of bands of frequencies should be set aside
}internationally for radio astronomy\footnote{These bands should lie at approximately every octave above 30 MHz
and should have bandwidths of about 1 percent of the center frequency.}
\item that special international protection should be afforded to the \textbf{hydrogen
line} (1400-1427 MHz), the \textbf{hydroxyl (OH) lines} (1645-1675
MHz), and to the predicted \textbf{deuterium line} (322-329 MHz)...
\end{enumerate}
Since 1959 a large number of spectral lines from a wide variety of
atoms and molecules in space have been discovered, then the frequency
range of radio astronomy now extends to at least 500 GHz. In particular
frequencies of the CO molecule (at 115, 230, and 345 GHz), isotopes
(at 110, 220, and 330 GHz) and the maser of H2O at 22,235GHz {[}29{]},
are critical to many aspects of astronomy, see also {[}21{]}. 

Radio astronomers have been engaged for decades in the work of the
United Nations Agency ITU to regulate the international use of the
radio frequency spectrum. Their efforts ensured a limited number of
narrow bands of the spectrum received protection to allow radio astronomy
to develop and conduct essential and unique research.

Despite the special international protection for Radio-astronomy,
some sources of radio frequency interference (RFI) are inescapable.
While radio astronomers can minimize the effects of many terrestrial
sources by placing their telescopes at remote sites, none can escape
from RFI generated by satellite transmitters, such as those of the
Iridium System, SpaceX, and others.

Whilst there is legislation in place where radio observatories are
placed (e.g. at the two SKA sites in Australia and South Africa) to
protect the telescopes from ground-based radio interference at those
frequencies, the use of air and space-borne radio communications is
regulated on a collaborative international basis.

What is not widely acknowledged is that the development of the latest
generation telecommunication networks (both from space and from Earth)
already has a profound impact on radio-astronomical observations (at
all sub-bands): with LEO satellite fleets it is quite sure that the
situation could become unbearable.
\begin{figure*}[t]
\includegraphics[width=0.8\paperwidth,height=0.4\paperwidth]{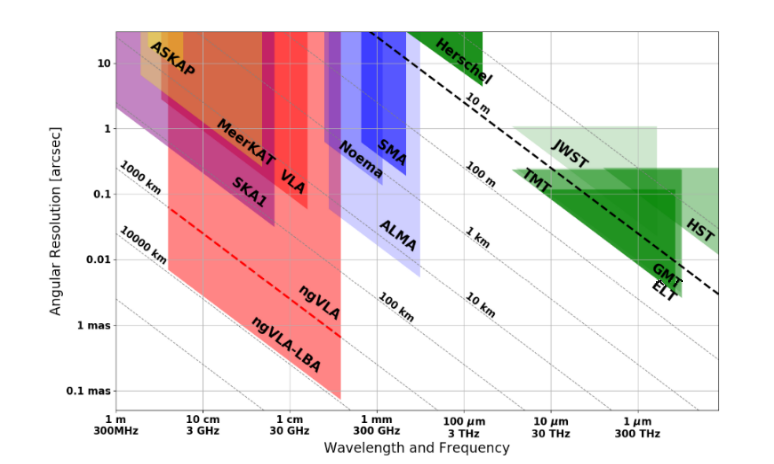}

\caption{Spatial resolution versus frequency set by the maximum baseline of
the ngVLA compared to other existing and planned facilities, see {[}18{]}.}
\end{figure*}

In particular, low Earth orbit satellite\textquoteright s spectral
windows identified to communicate with earth stations in the Ku (12-18GHz),
Ka (27-40GHz) and V (40-75GHz) bands will overlap with the nominal
radio-astronomy bands and so will interfere with ground radio telescopes
and radio interferometers, making the radio detectors enter in a non-linear
regime in the K band (18-26.5GHz) and in Q band (33-50GHz). This fact
will irreparably compromise the whole chain of analysis in those bands
with repercussions on our understanding of the Universe, or even,
making the astrophysics community blind to these spectral windows
from the ground, see Fig 6.

There are different projects in development for ground based radio-astronomy
that will significantly overlap with telecommunication signals coming
from the satellites' constellations in orbit: 
\begin{itemize}
\item The \textbf{Next Generation Very Large Array, ngVLA} and \textbf{ngVLA
Long Baseline Array, LBA} {[}18{]}: located in New Mexico, west Texas,
Arizona, and northern Mexico. The VLA will use 6 radio-bands: 2,4GHz,
8GHz, 16GHz, 27GHz, 41GHz and 93GHz. 
\item The \textbf{Square Kilometer Array, SKA} {[}19{]}, {[}23{]} will interfere
with Ku satellites communication bands.
\item The \textbf{Atacama Large Millimeter Array, ALMA} {[}22{]}, the world-leading
mm and sub-mm observatory built in Atacama, Chile, with enormous expenses
spent by a broad international community, facility that has brought
us many significant discoveries and played a crucial role in the global
system of EHT (first image of BH ever, published in April 2019), has
its Bands 1, and 2+3 exactly in the potentially polluted part of the
spectrum. 
\end{itemize}
To aggravate the matter, with the current technological development,
the planned density of radio frequency transmitters is impossible
to envisage. In addition to millions of new commercial wireless hot
spot base stations on Earth directly connected to the about 50,000
new satellites in space, will produce at least 200 billion of new
transmitting objects, according to estimates, as part of the Internet
of Things (IoT) by 2020-2022, and one trillion of objects a few years
later. 

Such a large number of radio-emitting objects could make radio astronomy
from ground stations impossible without a real protection made by
countries\textquoteright{} safe zones where radio astronomy facilities
are placed. 

We wish to avoid that technological development without serious control
turns radio astronomy practice into an ancient extinct science.

\section{How to protect the Astronomical Sky?}

To answer this question, we must remember some International Conventions
and Treaties.

The Preamble of the \textbf{World Heritage Convention} holds that
\textquotedblleft the deterioration or disappearance of any item of
the cultural or natural heritage constitutes a harmful impoverishment
of the heritage of all the nations of the world\textquotedblright{}
This protection appears again in the 1994 Universal Declaration of
Human Rights for Future Generations:\\
\fbox{\begin{minipage}[t]{1\columnwidth - 2\fboxsep - 2\fboxrule}%
\noun{Persons belonging to future generations have the right to an
uncontaminated and undamaged Earth, including pure skies; they are
entitled to its enjoyment as the ground of human history of culture
and social bonds that make each generation and individual a member
of one human family.}%
\end{minipage}}

The \textbf{UNESCO} has undertaken activities for the safeguarding
of cultural heritage related to astronomy under the \textquotedblleft Astronomy
and World Heritage\textquotedblright{} project launched by the World
Heritage Centre in 2003. This concept was taken up again by UNESCO
in 2005 as:\noun{}\\
\noun{}%
\fbox{\begin{minipage}[t]{1\columnwidth - 2\fboxsep - 2\fboxrule}%
\noun{The sky, our common and universal heritage, is an integral part
of the environment perceived by humanity. Humankind has always observed
the sky either to interpret it or to understand the physical laws
that govern the universe. This interest in astronomy has had profound
implications for science, philosophy, religion, culture and our general
conception of the universe.}%
\end{minipage}}

This in turn led to the following concepts:\noun{}\\
\noun{}%
\fbox{\begin{minipage}[t]{1\columnwidth - 2\fboxsep - 2\fboxrule}%
\noun{astronomical observations have profound implications for the
development of science, philosophy, religion, culture and the general
conception of the universe\dots{} discoveries of astronomers in the
field of science have had an influence not only on our understanding
of the universe but also on technology, mathematics, physics and social
development in general\dots{} the cultural impact of astronomy has
been marginalized and confined to a specialized public.}%
\end{minipage}}

These protections for Starlight are necessary as the impact that Starlight
has held on humanity has been expressed in works of religion, art,
literature, science, philosophy, business, and travel.

Enforcement of the Right to Starlight:\noun{}\\
\noun{}%
\fbox{\begin{minipage}[t]{1\columnwidth - 2\fboxsep - 2\fboxrule}%
\noun{International law enforces international legal obligations,
including property interests. Here, World Heritage is the property
of all humankind, and while there may be protective laws, enforcing
this is another matter, as only States can sue other States under
this type of international treaty. }\textbf{\noun{\uline{A State
is responsible for the activities that occur within its jurisdiction
-- whether they are authorized or unauthorized}}}\noun{.}%
\end{minipage}} 

Thus:\noun{}\\
\noun{}%
\fbox{\begin{minipage}[t]{1\columnwidth - 2\fboxsep - 2\fboxrule}%
\noun{Within the framework of International Law and State based legal
instruments, Protection of Starlight could then be implemented in
the same manner:}\\
1. \noun{Reaffirms the sovereign rights and responsibilities, towards
the International Community, of each State for the protection of its
own cultural and natural heritage;}\\
\noun{2. Calls upon the International Community to provide all the
possible assistance needed to protect and conserve the cultural and
natural heritage of Starlight;}\\
\noun{3. Invites the authorities of States to take appropriate measures
in order to safeguard the cultural and natural heritage of Starlight;}\\
\noun{4. Further invites the States to co-operate with UNESCO, the
World Heritage Committee, the UNWTO, and the Starlight Initiative
with a view to ensuring effective protection of its cultural and natural
heritage in Starlight.}%
\end{minipage}}

Having established these rights under international law, the conclusion
is that there exist duties for both States and international organizations
to protect the World Heritage Right to Starlight, as well as, their
duties to foster the rights of travelers, hosts, and providers of
travel to enjoy this Starlight \textquotedblleft property interest\textquotedblright{}
that belongs to all humanity. The existing legal instruments demonstrate
the protection for the Right to Starlight, but it is the States that
act as custodians of World Heritage that are charged with ensuring
these rights are enforceable, and in turn made available to all of
humanity.

\subsection{On the legal side}

SpaceX private company has received permission from many USA government
agencies (e.g. Federal Communication Commission, FCC) to launch these
satellites into orbit. So there could be a legal claim, within the
US legal system, to halt the progress of Starlink.

Also, as it turns out, according to the Outer Space Treaty and its
progeny, there are no private companies operating in outer space,
but only governments can operate in outer space. And the legal process
is that the state government, this time the USA government, is legally
responsible for all objects sent into outer space that launch from
USA borders. That means, that it is the USA government that is responsible
for the harm caused by its corporation, Starlink, sending objects
into orbit that cause harm. 

So under this international law, any country that suffers harm by
Starlink can sue the United States government in the International
Court of Justice in the Hague. The harm here is damage to our cultural
heritage, the night sky, and monetary damages due to the loss of radio
and other types of astronomy. For the scientists, the owners of the
observatories have a legal argument that they have and will continue
to lose money spent for their research based on Earth based observatories.
Furthermore, Universities that own the observatories are state owned
universities, so it is the government that owns the observatories
that have lost financially because of their interruption of study
of the night skies.

So it is essential that a government, like Chile, Italy or France,
sues the USA in the International Court of Justice.

\textbf{If no national or international entity will stop this displacement
}\textbf{\uline{the right of the private company SpaceX will become
acquired at the beginning of March 2020}}.

How should the international astronomical community mobilize in order
to stop further Starlink launches?
\begin{enumerate}
\item Sue in court for luminous pollution not taken into account by US FCC:
The FCC\textquoteright s lack of review of these commercial satellite
projects violates the National Environmental Policy Act, NEPA, which
obligates all federal agencies to consider the environmental impacts
of any projects they approve. So in the most basic sense, SpaceX\textquoteright s
satellites displacement authorization would be unlawful, see {[}31{]}.
\item Sue in court for lack of jurisdiction and jurisprudence of US FCC
to authorize private not geostationary satellites over other states
and nations.
\item Sue in the International Court of Justice, ICJ the USA government
to put on hold further Starlink launches to quantify the loss of public
finances in damaging national and international astronomical projects.\footnote{Though there are no international law that restrict mega constellations,
to deploy and dispatch mega constellations an international agreement
among states is needed, since satellites can not be located only over
a single state (e.g. USA) but, being in LEO, they move around the
globe passing over different states/nations/continents. This is a
lack of jurisdiction of FCC authorization. In particular the International
Court of Justice, ICJ, can be called into question whenever there
is a dispute of international jurisdiction or between member states
of the United Nations on the basis of international norms, treaties
and / or their violations. In the beginning of chapter 5 it was explained
how the World Heritage Convention regarding the ``right of night
sky / starlight'' belongs to universal human rights and so no state
can decide to contravene this convention if it interferes with the
enjoyment of that right for other states. The pretext for appealing
to the United Nations and the International Court of Justice (ICJ)
is the loss of scientific value of the investments made for ground
based projects by each state (damaged by SpaceX). Each damaged state,
being damaged as conseguence of a violation for an international treaty,
the issue cannot be settled with a simple money compensation, but
with an inhibition of the damage before the same occurs (and not after).} 
\end{enumerate}

\subsection{From the astronomers' side}

An \textbf{international appeal/petition from astronomers} was launched
in January 2020 and, at the time of writing, thousands of astronomers
involved with astronomical observatories and facilities, have subscribed
the appeal, see {[}25{]}. Another \textbf{open letter} has been prepared
reguarding same concerns on the satellites constellations deployment
for the further space missions and to raise awareness to US Senate,
and US commisisons on the possibility that occurs the Kessler Syndrome,
which is a realistic scenario with all these orbiting objects, see
{[}32{]}. 

Requests from the astronomical community to governments, institutions,
and agencies all around the world are:
\begin{enumerate}
\item to be committed to provide legal protection to ground astronomical
facilities in all of the available observation electromagnetic windows. 
\item to put on hold further Starlink launches (and other projects) and
carry out an accurate moratorium on all technologies that can negatively
impact astronomical space based and ground based observations, or
impact on the scientific, technological and economic investments that
each State engages in astrophysical projects. 
\item to put in place a clear evaluation of risks and predictive impacts
on astronomical observatories (i.e. loss of scientific and economic
value), giving stringent guidelines to private individuals, societies
and industries to plan satellite investments without clearly understanding
all of the negative effects on outstanding astronomical facilities. 
\item that the US Federal Communications Commission (FCC) and any other
national agency be wary of granting permission to ship non-geostationary
low-orbit satellites into orbit or alternatively to limit the authorization
of only satellites being above the airspace of the \textquotedblleft home
country\textquotedblright . 
\item to demand a worldwide orchestration, where national and international
astronomical agencies can impose the right of veto on all those projects
that negatively interfere with astronomical outstanding facilities. 
\item to limit and regulate the number of telecommunication satellite fleets
to the \textquotedblleft strictly necessary number\textquotedblright{}
and to put them in orbit only when old-outdated technology satellites
are deorbited, according to the Outer Space Treaty (1967) - the Art
IX {[}8{]}, and the United Nations Guidelines for the Long-term Sustainability
of Outer Space Activities (2018) -- guideline 2.2(c) {[}9{]}, requiring
the use of outer space be conducted \textquotedblleft so as to avoid
{[}its{]} harmful contamination and also adverse changes in the environment
of the Earth\textquotedblright{} and {[}\dots omissis...{]} risks
to people, property, public health and the environment associated
with the launch, in-orbit operation and re-entry of space objects\textquotedblright .
\end{enumerate}

\section{Conclusions }

Avoiding damages in astronomical ground based observations arised
from the displacement of satellite constellations is absolutely mandatory
for safeguarding not only the economic and scientific investment,
committed by international institutions and single nations, but also
to continue efficiently to monitor possible impact events to guard
and alert the humankind. 

All of these requests come from the heartfelt concern of scientists
arising from threatens to be barred from accessing the full knowledge
of the Cosmos and the loss of an intangible asset of immeasurable
value for humanity. In this context it is absolutely necessary to
put in place all possible measures to protect the night sky right
also on the legal side as stated by the Universal Declaration of Human
Rights for Future Generations. 

To ensure this safeguard action it would be desirable to adopt contingent
and limiting resolutions to be ratified as \uline{shared international
rules}, which must be adopted by all space agencies to ensure protection
for astronomical bands observable from the ground. All of this to
continue to admire and study our Universe, for as long as possible.

In the meanwhile \uline{all private displacement of satellites
constellation project must be put on hold}; every ``national'' authorization
to launch not geostationary LEO satellite fleets must be withdrawn
and avoided as well.

\textquotedbl One person's freedom ends where another's begins\textquotedbl .
The public authorities (of every state) are entitled (and obliged)
to enforce regulations, which take care that the above statement comes
true. The right to see the sky in natural state belongs to our rights
and freedoms alike the right to breath unpolluted air, drink clean
water or sleep in a quiet environment during the night. 

$ $

\begin{doublespace}
\texttt{\textbf{\noun{\large{}References}}}{\large\par}
\end{doublespace}

\texttt{{[}1{]} International Astronomical Union, IAU statements: \href{https://www.iau.org/news/announcements/detail/ann19035/?lang}{www.iau.org/ann19035}}
\begin{flushleft}
\texttt{{[}2{]} ESO Very Large Telescope: \href{https://www.eso.org/public/usa/teles-instr/paranal-observatory/vlt/?lang}{www.eso.org/paranal-observatory/vlt}}
\par\end{flushleft}

\begin{flushleft}
\texttt{{[}3{]} Large Binocular Telescope, LBT: \href{http://www.lbto.org/}{http://www.lbto.org/}}
\par\end{flushleft}

\begin{flushleft}
\texttt{{[}4{]} ESO Extremely Large Telescope, E-ELT: \href{https://www.eso.org/public/usa/teles-instr/elt/?lang}{www.eso.org/elt}}
\par\end{flushleft}

\begin{flushleft}
\texttt{{[}5{]} Large Syhoptic Survey Telescop, LSST: \href{https://www.lsst.org\%20\%E2\%80\%93\%20https://en.wikipedia.org/wiki/Vera_C._Rubin_Observatory}{Vera\_C.\_Rubin\_Observatory}}
\par\end{flushleft}

\begin{flushleft}
\texttt{{[}6{]} ESO VLT Survey Telescope: \href{https://www.eso.org/public/italy/teles-instr/paranal-observatory/surveytelescopes/vst/}{surveytelescopes/vst}}
\par\end{flushleft}

\begin{flushleft}
\texttt{{[}7{]} Pan-STARRS Telescope: \href{https://panstarrs.stsci.edu/}{panstarrs.stsci.edu}}
\par\end{flushleft}

\begin{flushleft}
\texttt{{[}8{]} United Nations Office for Outer Space Affairs, ``Treaty
on Principles Governing the Activities of States in the Exploration
and Use of Outer Space, including the Moon and Other Celestial Bodies'': \href{https://www.unoosa.org/oosa/en/ourwork/spacelaw/treaties/introouterspacetreaty.html}{introouterspacetreaty.html}}
\par\end{flushleft}

\begin{flushleft}
\texttt{{[}9{]} Committee on the Peaceful Uses of Outer Space, ``Guidelines
for the Long-term Sustainability of Outer Space Activities '': \href{https://www.unoosa.org/res/oosadoc/data/documents/2018/aac_1052018crp/aac_1052018crp_20_0_html/AC105_2018_CRP20E.pdf}{https://www.unoosa.org}}
\par\end{flushleft}

\begin{flushleft}
\texttt{{[}10{]} Simulated prediction of \textquotedblleft only\textquotedblright{}
12k Starlink satellites positions in the sky: \href{https://youtu.be/LGBuk2BTvJE}{www.youtu.be/LGBuk2BTvJE} }
\par\end{flushleft}

\begin{flushleft}
\texttt{{[}11{]} Simulated prediction of \textquotedblleft only\textquotedblright{}
12k Starlink satellites naked eye view of the sky: \href{https://www.youtube.com/watch?v=z9hQfKd9kfA}{www.youtube.com/9hQfKd9kfA}}
\par\end{flushleft}

\begin{flushleft}
\texttt{{[}12{]} Visualization tool to find, plot and search satellite
orbits: \href{https://celestrak.com/cesium/orbit-viz.php}{https://celestrak.com/orbit-viz}}
\par\end{flushleft}

\begin{flushleft}
\texttt{{[}13{]} Phil Cameron, ``The right to Starlight Under International
Law''}
\par\end{flushleft}

\begin{flushleft}
\texttt{{[}14{]} Starlight Initiative ``Declaration in defense of
the night sky and the right to starlight'': \href{https://issuu.com/pubcipriano/docs/starlightdeclarationen}{starlightdeclaration}}
\par\end{flushleft}

\begin{flushleft}
\texttt{{[}15{]} Marian Cipriano, ``Starlight: a common heritage'',
Published online by Cambridge University Press: 29 June 2011 }
\par\end{flushleft}

\begin{flushleft}
\texttt{{[}16{]} Marin, C., and Jafar, J. (eds) 2008,Starlight: A
common Heritage(Tenerife: Instituto de As-trof\'{ }\i sica de Canarias)}
\par\end{flushleft}

\begin{flushleft}
\texttt{{[}17{]} Starlight Scientific Committee Report 2009,Starlight
Reserve Concept: \href{http://www.starlight2007.net/pdf/StarlightReserve.pdf}{http://www.starlight2007.net}}
\par\end{flushleft}

\begin{flushleft}
\texttt{{[}18{]} A.J. Beasley, E. Murphy, R. Selina, M. McKinnon \&
the ngVLA Project Team, ``The Next Generation Very Large Array (ngVLA)''
NRAO}
\par\end{flushleft}

\begin{flushleft}
\texttt{A citation: \cite{book-full}}
\par\end{flushleft}

\begin{flushleft}
\texttt{{[}19{]} Square Kilometer Array, SKA: \href{https://www.skatelescope.org/}{https://www.skatelescope.org}}
\par\end{flushleft}

\begin{flushleft}
\texttt{{[}20{]} Avinash A. Deshpande and B. M. Lewis, ``Iridium
Satellite Signals: A Case Study in Interference Characterization and
Mitigation for Radio Astronomy Observations'', DOI: 10.1142/S225117171940009,
arXiv:1904.00502 {[}astro-ph.IM{]}}
\par\end{flushleft}

\begin{flushleft}
\texttt{{[}21{]} Committee on Radio Frequencies Board on Physics and
Astronomy Commission on Physical Sciences, Mathematics, and Applications
National Research Council ``VIEWS OF THE COMMITTEE ON RADIO FREQUENCIES
CONCERNING FREQUENCY ALLOCATIONS FOR THE PASSIVE SERVICES AT THE 1992
WORLD ADMINISTRATIVE RADIO CONFERENCE `` }
\par\end{flushleft}

\begin{flushleft}
\texttt{{[}22{]} Atacama Large Millimeter Array, ALMA: \href{https://www.eso.org/public/usa/teles-instr/alma/?lang}{https://www.eso.org}}
\par\end{flushleft}

\begin{flushleft}
\texttt{{[}23{]} SKA statements on satellites contellations: \href{https://www.skatelescope.org/news/ska-statement-on-satellite-constellations/}{ska-statement-on-satellite-constellations}}
\par\end{flushleft}

\begin{flushleft}
\texttt{{[}24{]} PATRICIA COOPER, ``SATELLITE GOVERNMENT AFFAIRS
SPACE EXPLORATION TECHNOLOGIES CORP. (SPACEX)'': \href{https://www.commerce.senate.gov/services/files/6c08b6c2-fe74-4500-ae1d-a801f53fd279}{https://www.commerce.senate.gov}}
\par\end{flushleft}

{[}25{]} APPEAL BY ASTRONOMERS: \href{https://astronomersappeal.wordpress.com}{https://astronomersappeal.wordpress.com}\\
{[}26{]} Starlink testing encrypted internet form military US Air
Force purposes: \href{https://www.reuters.com/article/us-spacex-starlink-airforce/musks-satellite-project-testing-encrypted-internet-with-military-planes-idUSKBN1X12KM}{https://www.reuters.com/article}\\
{[}27{]} Starlink APPLICATION FOR APPROVAL FOR ORBITAL DEPLOYMENT
AND OPERATING AUTHORITY FOR THE SPACEX NGSO SATELLITE SYSTEM: \href{https://cdn.arstechnica.net/wp-content/uploads/2017/05/Legal-Narrative.pdf\%20}{Legal-Narrative.pdf }\\
{[}28{]} Starlink simulations from deepskywatch: \href{http://www.deepskywatch.com/Articles/Starlink-sky-simulation.html}{http://www.deepskywatch.com}\\
{[}29{]} Volvach et al 2019, ``An Unusually Powerful Water-Maser
Flare in the Galactic Source W49N'', doi: \href{https://ui.adsabs.harvard.edu/link_gateway/2019ARep...63..652V/doi:10.1134/S1063772919080067}{2019ARep...63..652V/doi:10.1134}

{[}30{]} Donald J. Kessler, Burton G. Cour-Palais, ``Collision frequency
of artificial satellites: The creation of a debris belt'', doi: \href{https://doi.org/10.1029/JA083iA06p02637}{https://doi.org/10.1029/JA083iA06p02637}

{[}31{]} Ramon J. Ryan, Note, The Fault In Our Stars: Challenging
the FCC\textquoteright s Treatment of Commercial Satellites as Categorically
Excluded From Review Under the National Environmental Policy Act,
22 VAND. J. ENT. \& TECH. L. (forthcoming May 2020)

{[}32{]} David Dubois, NASA ``Open Letter to FCC, US Senate and Commissions
and SpeceX``, \href{https://docs.google.com/document/d/1fScGypCWEG6j2S7oPBb8u0ibQ5lRA}{OpenLEtterforFCCandNASA} 
\end{document}